# Statistical Study of Visual Binaries

H. I. Abdel-Rahman, M. I. Nouh and W. H. Elsanhoury

Astronomy Dept., National Research Institute of Astronomy and Geophysics, 11421, Helwan, Cairo, Egypt.

**Abstract:** In this paper, some statistical distributions of wide pairs included in Double Star Catalogue are investigated. Frequency distributions and testing hypothesis are derived for some basic parameters of visual binaries. The results reached indicate that, it was found that the magnitude difference is distributed exponentially, which means that the majority of the component of the selected systems is of the same spectral type. The distribution of the mass ratios is concentrated about 0.7 which agree with Salpeter mass function. The distribution of the linear separation appears to be exponentially, which contradict with previous studies for close binaries.

**1. Introduction**

Investigation of visual binaries is a part of studying the physical properties of the whole family of double stars. Some statistical properties of these stars, such as the distribution function of their linear separation, leads to important information related to their origin as well as, in some cases to their subsequent evolution. A further difficulty in discovering visual binaries is created for the very distant companions. Such companions are less likely to form a genuine binary system with the primary star, being rather, optical companions (Halbwachs, 1983).

To avoid the including of the optical pairs in the statistics of binary stars, a particular spatial separation of a pair of stars is somewhat arbitrarily defined as the maximum possible for a real binary system. Some genuine binaries are then undoubtedly omitted from the statistics while some optical pairs are incorrectly included. If the limit is carefully chosen, however, the probable total number of the optical pairs is negligible. From the assumed upper limit of the real separation of the stars, a statistical relation between the angular separation and the apparent magnitude of the binary system can be derived and used to test the binary nature of the any pair of stars, Abt (1986, 1988), Halbwach (1983, 1986), Nouh and Sharaf (2012).

In the present paper, we shall investigate some statistical distributions (e.g. magnitude difference, mass ratios, linear separations and the luminosity function) and examine the relations between them.

## 2. Data and Method of Analysis

We have performed a statistical analysis of visual binaries from the Washington Double Star Catalogue (WDS), Mason et al. (2001). The number of stars that may produce optical pairs may be reduced by adding a condition that should be satisfied by the components of physical binaries. Heintz's (1978) criterion says that only pairs within certain ranges of $\Delta m$ and separation will have been identified as binaries, his criterion was stated as

$$C = 0.22 \, \Delta m - \log \rho'' \leq 0.5 \qquad (1)$$

for $m_a \leq 9.5$ where $m_a$ is the apparent magnitude of the primary.

The entries in the WDS catalogue were incomplete in some cases, so, the following procedure was applied.

- Systems that having separation values in the catalogue only chosen for this study.
- The primary component of luminosity class V is only accepted.

Following this procedure, we get 2837 systems with luminosity class V primaries. For the selected systems we had determined the masses and mass ratios q of components making use of Sp-$M_V$ and Sp-mass (Sp denotes the spectral type) relations by Allen (1973). The distance to star $d$ was determined by using the relation

$$m - M_v = 5 \log d - 5 \qquad (2)$$

The true, a", and projected separations, $\rho''$, are related on average as (Couteau, 1981),

$$a'' = 1.25 \, \rho''. \qquad (3)$$

The linear separation could be computed by

$$a(\text{AU}) = a'' \, d(\text{pc})$$

## 3. Results

### 3.1. Frequency Distribution of $\Delta m_a, m_a$, q, a, $\rho''$

The frequency distribution of the physical parameters could be constructed as follows:

i- We determine the minimum and maximum values in the data and we get the range, R = maximum value – minimum value.
ii- Using Storges rule to determine the number of intervals (n).

$$n = 1 + 3.3 \log(N) \qquad (4)$$

iii- The length of intervals is given by L where $L = \frac{R}{n}$, (5)

where N is the number of data.

### 3.3.1. The Apparent Magnitude Difference

After examining the data, we found that the large part of the data (99%) for the apparent magnitude difference (Δm) lie between 0 to 7.8 and the number of binaries become 2816 after deleting the anomalous points.

Table 1 contains some descriptive statistics for Δm such as N (number of data, $\bar{X}$ (mean), $\sigma$ (standard deviation), minimum and maximum values and the range.

Table 1: Some Descriptive statistics for Δm

| N (no. of data) | mean | Standard deviation | Minimum value | Maximum value | Range |
|---|---|---|---|---|---|
| 2816 | 2.014 | 1.75 | 0 | 7.8 | 7.8 |

We apply Equations (4) and (5) and the information in Table 1; we get the number of intervals (n) and the length of interval (L) using n = 12 and $L = 0.65$ and then we construct Table 2, where the first column is the intervals (classes), the second is the center of the interval ( average of the low and upper limits) and the third column is the frequencies (the numbers corresponds each interval). We note in the table that, the binaries are concentrated at the interval 0 - 0.65 and the frequency is 817, i.e. 29 % from the whole sample. Generally, the magnitude difference of binaries is concentrated between the intervals 0 -5.2 with 2644 binaries, i. e. approximately 94 % and 6 % are concentrated in the interval 5.2 -7.8.

Figure 1 shows that the frequency distribution of the magnitude difference of binaries and it is distributed exponentially. The intrinsic distribution of visual binaries with respect to the magnitude difference reflects their distribution with respect to the mass ratios $q$. From the graph we can notice that the majority of the wide systems are approximately concentrated about $\Delta m \simeq 0$, this means that the components of these systems are almost of the same spectral type. The distributions of the difference apparent magnitude, the linear separation and the projected separation of binaries distributed exponentially.

Table 2: Frequency distribution of the magnitude difference.

| intervals (classes) | Center of sets | Frequencies (number) |
| --- | --- | --- |
| 0 - 0.65 | 0.325 | 817 |
| 0.65 – 1.3 | 0.98 | 467 |
| 1.3 – 1.95 | 1.625 | 302 |
| 1.95 – 2.6 | 2.275 | 349 |
| 2.6 – 3.25 | 2.925 | 243 |
| 3.25 – 3.9 | 3.575 | 204 |
| 3.9 – 4.55 | k4.225 | 172 |
| 4.55 – 5.2 | 4.875 | 90 |
| 5.2 – 5.85 | 5.525 | 57 |
| 5.85 – 6.5 | 6.175 | 54 |
| 6.5 – 7.15 | 6.825 | 31 |
| 7.15 – 7.8 | 7.475 | 30 |
| Summation |  | 2816 |

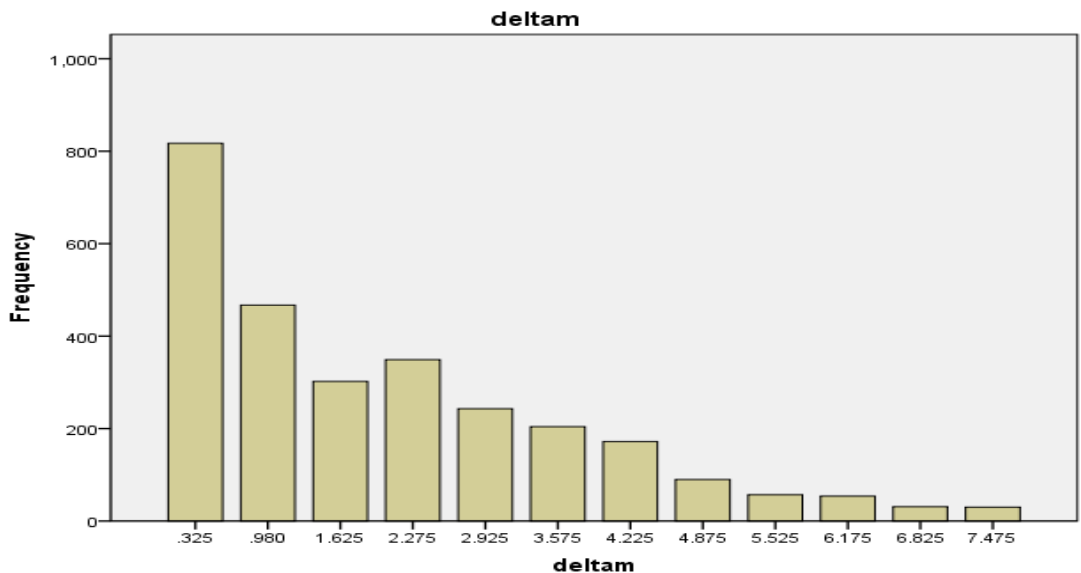

Fig 1. Distribution of the magnitude difference of Wide visual binaries.

### 3.1.2 The Apparent Magnitude

Table 3 contains the basic information which we use to construct the frequency distribution for the apparent magnitude $m_a$.

Table 3: Some descriptive parameters of $m_v$

| N | mean | Standard deviation | Minimum value | Maximum value | Range |
|---|---|---|---|---|---|
| 2839 | 8.61 | 1.51 | 1.3 | 12.5 | 11.2 |

After we applied Equations (4) and (5), we get the number of intervals n= 12 and the length of interval L = 0.93 and then the frequency distribution is listed in Table 4.

In Table 4, the larger frequencies are 872 binaries at the interval 8.74 - 9.67 and 662 at the interval 9.67 -10.6, which represent about 54 % of all binaries. We conclude that the frequency distribution of the primary apparent magnitude $m_v$ is approximately exponential growth, Figure 2.

Table 4: the distribution of the apparent magnitude $m_a$

| Sets (classes) | Center of sets | Frequencies (number) |
|---|---|---|
| 1.3 – 2.23 | 1.765 | 2 |
| 2.23 – 3.16 | 2.695 | 4 |
| 3.16 – 4.09 | 3.625 | 13 |
| 4.09 – 5.02 | 4.555 | 63 |
| 5.02 – 5.95 | 5.485 | 119 |
| 5.95 – 6.88 | 6.415 | 192 |
| 6.88 – 7.81 | 7.345 | 336 |
| 7.81 – 8.74 | 8.275 | 486 |
| 8.74 – 9.67 | 9.205 | 872 |
| 9.67 – 10.6 | 10.135 | 662 |
| 10.6 – 11.53 | 11.065 | 83 |
| 11.53 -12.5 | 12.015 | 7 |
| Summation |  | 2839 |

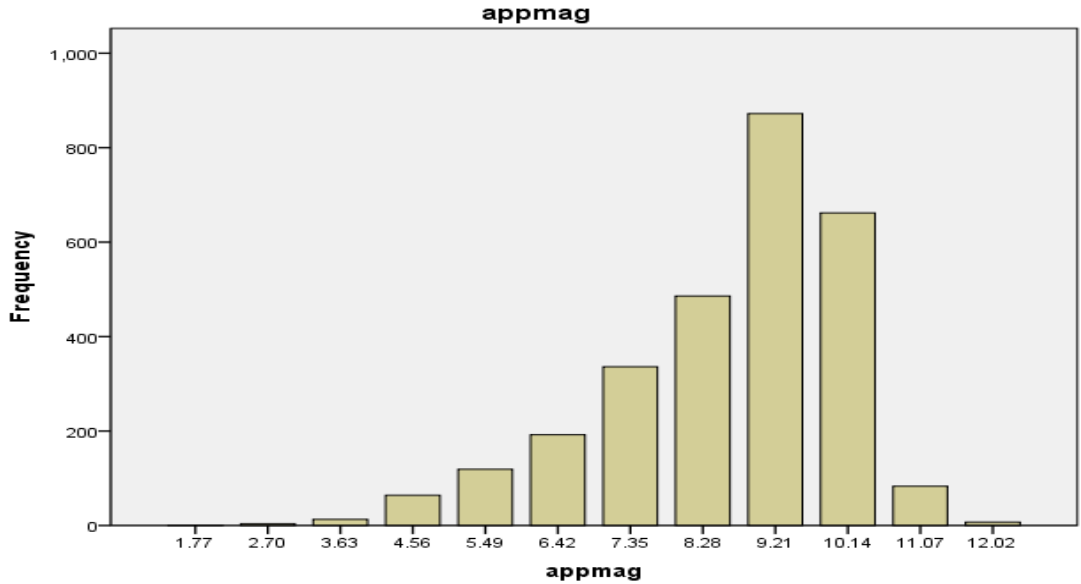

Fig 2. Frequency distribution of the primary apparent magnitude $m_a$ of visual binaries.

### 3.1.3 Mass Ratio

The basic statistics for the mass ratio are tabulated in Table 5 and the frequency distribution is in Table 6. In Table 6, we note that the arrangement of higher frequencies (concentration) are 385 binaries concentrated in the interval (0.826 - 0.911), 342 at interval (0.911 - 0.966), 330 binaries at (0.741 – 0.826) respectively. The total binaries in the last three intervals are 1057 = 37 % from all binaries concentration and 2424 binaries are concentrated in the intervals from 0.401 to 1.081 this means that 85.4% and 96% from the mass ratio for binaries concentrated between 0.231 to 1.081 and 4 % between 0.061 to 0.231. Figure 3 shows that the distribution of the mass ratio appears to be linear.

Table 5: The basic statistics for the mass ratio (q)

| N (no. of data) | mean | Standard deviation | Minimum value | Maximum value | Range |
|---|---|---|---|---|---|
| 2839 | 0.67 | 0.23 | 0.061 | 1.078 | 1.017 |

Table 6: The frequency distribution of mass ratio

| Sets (classes) | Center of sets | Frequencies |
|---|---|---|
| 0.061 – 0.146 | 0.1035 | 21 |
| 0.146 – 0.231 | 0.1885 | 78 |
| 0.231 – 0.316 | 0.2735 | 124 |
| 0.316 – 0.401 | 0.3585 | 192 |
| 0.401 – 0.486 | 0.4435 | 253 |
| 0.486 – 0.571 | 0.5285 | 323 |
| 0.571 – 0.656 | 0.6135 | 308 |
| 0.656 – 0.741 | 0.6985 | 287 |
| 0.741 – 0.826 | 0.7835 | 330 |
| 0.826 – 0.911 | 0.8685 | 385 |
| 0.911 – 0.996 | 0.9535 | 342 |
| 0.966 – 1.081 | 1.0385 | 196 |
| Summation | | 2839 |

The distribution of the mass ratios $q$ is tabulated in Table 6 and plotted in Figure 3. The maximum frequency of this distribution is concentrated about q=0.7. As stated by Vereshchagin et al. (1988) that, two selection factors affect the distribution of the mass ratios, the stellar magnitude, and the angular separation. Since we have in our study, wide systems and a stellar magnitude is almost less than $10^m$, therefore the effect of the two selection factors are decreased.

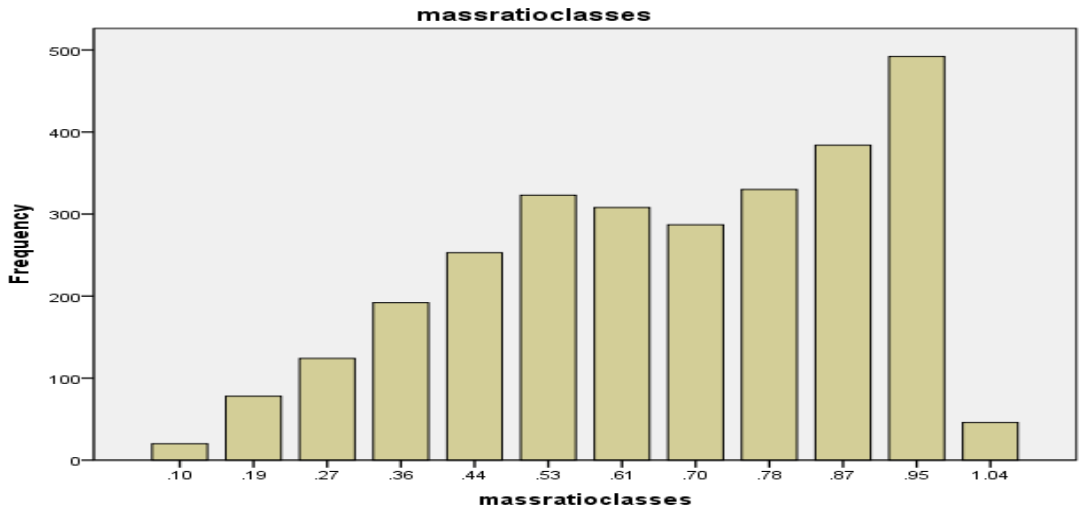

Fig 3. Frequency distribution of the mass ratio $q$ of Wide visual binaries.

Another factor which affects the distribution of $q$ is the number of the degenerate components which leads to misleading results. The proportion of the degenerate components $R$ can be determined with the aid of the two formulas (Halbwachs, 1984),

$$R = r/(1+r),$$

where

$$r(q) = q^{1.35} - q^{4.60},$$

for the case of the constant birth rate, and

$$r(q) = q^{1.35} - q^{7.85},$$

for linear birth rate. In this calculation the mass of primary must be larger than 1.045 $M_\odot$. The proportion $R$ is computed for both cases and plotted in Figure 6. From this curve we can conclude that the frequency of binaries with degenerate component is great when the distribution of binaries among mass ratios contains many systems with q near 0.7.

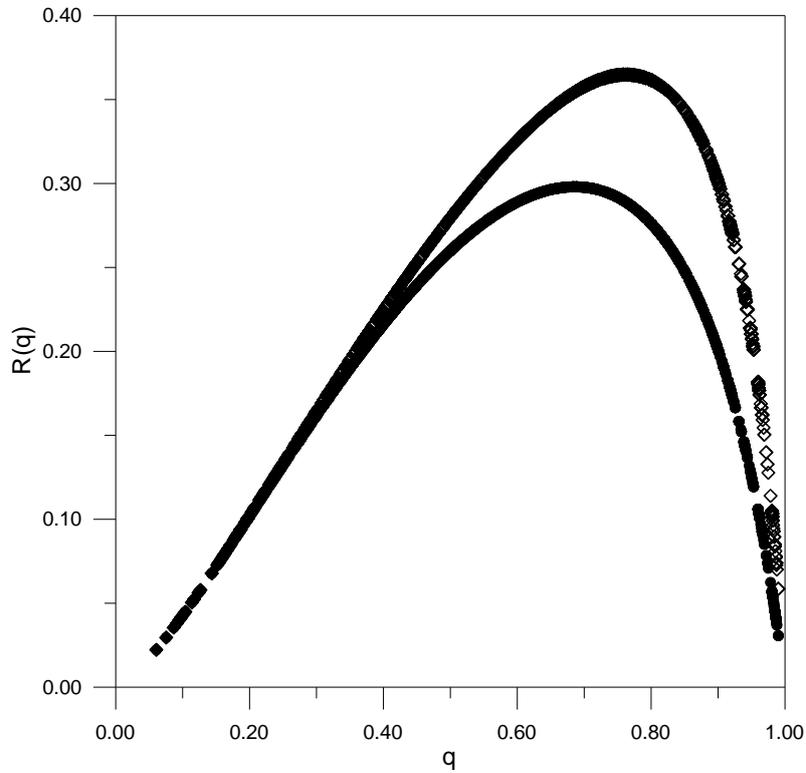

Fig 4. The proportion of wide pairs with one degenerate component *R*, related to the mass ratio *q*; lower curve: case of a constant birthrate; upper curve: linear decreasing birthrate reaching zero at present.

### 3.1.4 The Projected Separation

Firstly, we delete 834 anomalous points. After using the basic descriptive parameters in Table 7 and Equations (4) and (5), we get n = 12 and the length of interval = 190.

Table 7: the basic descriptive parameters

| N (no. Of data) | mean | Standard deviation | Minimum value | Maximum value | Range |
|---|---|---|---|---|---|
| 2005 | 641 | 580 | 6 | 2280 | 2273 |

In Table 8, the description is as in the above tables. We note that the frequencies are in descending order, this means that the big concentration of binaries at the interval 6- 196 is 56= 28% and the interval 196-386 is 370 = 18.5% … and so on. The concentration of binaries is located in the first 6 intervals about 80% and 20 % in the other 6 intervals. Finally, we conclude that the frequency distribution of projected separation for binaries is distributed exponentially as in Figure 5.

Table 8: The frequency distribution of projected separation

| Sets (classes) | Center of sets | Frequencies (number) | Percent % |
|---|---|---|---|
| 6 - 196 | 101 | 561 | 28.0 |
| 196 - 386 | 291 | 370 | 18.5 |
| 386 - 576 | 481 | 249 | 12.4 |
| 576 - 766 | 671 | 191 | 9.5 |
| 766 - 956 | 861 | 118 | 5.9 |
| 956 - 1146 | 1051 | 117 | 5.8 |
| 1146 - 1336 | 1241 | 93 | 4.6 |
| 1336 - 1526 | 1431 | 90 | 4.5 |
| 1526 - 1716 | 1621 | 70 | 3.5 |
| 1716 - 1906 | 1811 | 50 | 2.5 |
| 1906 - 2096 | 2001 | 51 | 2.5 |
| 2096 - 2286 | 2191 | 45 | 2.2 |
| Summation | | 2005 | 100.0 |

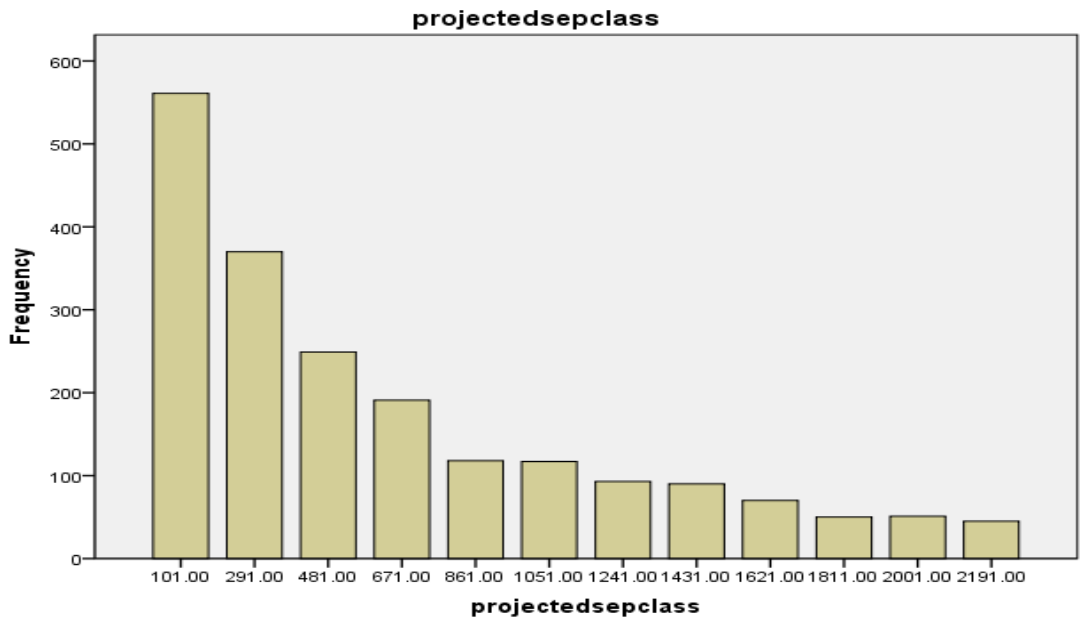

Fig 5. Frequency distribution of the projected separation of visual binaries.

### 3.1.5 The Linear Separation

After deleting 97 anomalous points and using the basic descriptive parameters in Table (9) and Equations (4) and (5), we get the number of intervals n= 12 and the length of each interval L= 1665.

Table 9: The Basic Parameters of the Linear Separation

| N (no. Of data) | mean | Standard deviation | Minimum value | Maximum value | Range |
|---|---|---|---|---|---|
| 2742 | 2561 | 3641 | 8 | 19992 | 19984 |

The frequency distribution of linear separation is illustrated in Table 10, in this table, we note that the largest frequencies of binaries are in interval 8 - 1673 1(700 binaries) about 62 % and 88% from binaries concentrated in the first 4 intervals from 8 to 6668 and 12 % in the others eighth intervals. Figure 5 show that the linear separation of binaries is distributed exponentially.

    The linear separations are another quantity which are tightly connected to the evolution of the wide pairs. Since we cannot determine the orbit of the wide binaries, Kuiper (1935), Couteau (1960), van Albada (1968) and Halbwachs (1983) stated that the linear separation is nearly equivalent to the semi-major axis of the orbit. This

distributions is shown in Figure 6. As appeared from the graph, the distributions are exponentially which contradict with the results derived for close binaries by Luyten (1967) and Halbwachs (1986). This contradiction may be attributed to the evolution processes in close binaries.

Table 10: the frequency distribution of linear separation

| Sets (classes) | Center of sets | Frequencies (number) | Percent % |
|---|---|---|---|
| 8 – 1673 | 840.5 | 1700 | 61.999 |
| 1673 – 3338 | 2505.5 | 378 | 13.786 |
| 3338 – 5003 | 4170.5 | 208 | 7.586 |
| 5003 – 6668 | 5835.5 | 126 | 4.595 |
| 6668 – 8333 | 7500.5 | 97 | 3.538 |
| 8333 – 9998 | 9165.5 | 75 | 2.735 |
| 9998 – 11663 | 10830.5 | 50 | 1.823 |
| 11663 - 13328 | 12495.5 | 31 | 1.131 |
| 13328 - 14993 | 14160.5 | 23 | 0.839 |
| 14993 – 16658 | 15825.5 | 20 | 0.729 |
| 16658 - 18323 | 17490.5 | 17 | 0.62 |
| 18323 - 19993 | 19158 | 17 | 0.62 |
| Summation | | 2742 | 100.0 |

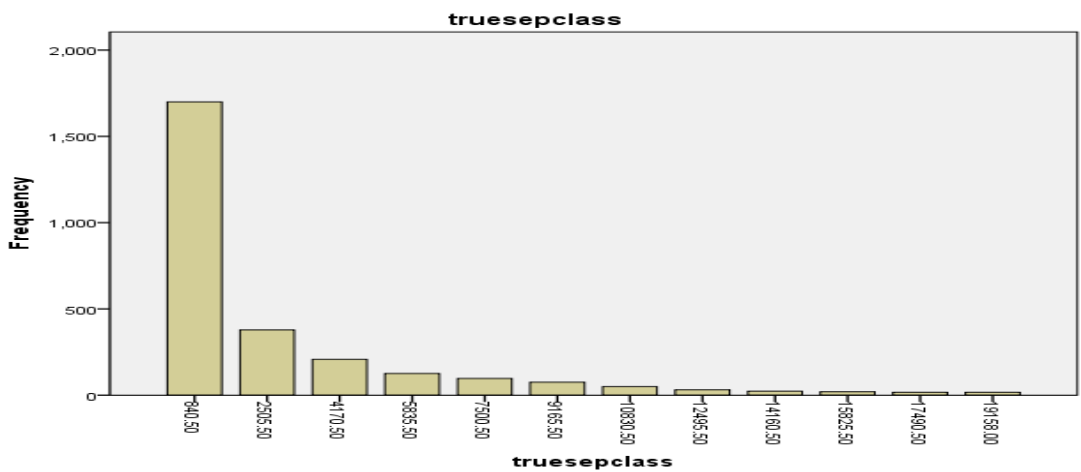

Fig 6. Frequency distribution of the linear separation of visual binaries.

## 3.2. Confidence Intervals and Testing Hypotheses

In this section, we do 95 % confidence interval (C. I.) and testing the hypothesis of populations means for the computed parameters.

Firstly, the 95 % C. I. for the population mean ($\mu$) is given by the following relation:

$$\bar{x} - d \leq \mu \leq \bar{x} + d \tag{6}$$

Where $\bar{x}$ is the sample mean and $y$ is the maximum error and is given by

$$y = Z_{\alpha/2} \sigma_{\bar{x}} \quad and \quad \sigma_{\bar{x}} = \frac{\sigma}{\sqrt{n}} \tag{7}$$

$\alpha = 0.05$, is the significant level and we get $Z_{\alpha/2}$ from the standard normal distribution table (Z-table).

Secondly, we test the hypothesis by the following procedure:

1. $H_0: \mu = \mu_0, \quad\quad H_1: \mu \neq \mu_0 \ or \ \mu > \mu_0 \ or \ \mu < \mu_0$ (8)

    Where $H_0 \ and \ H_1$ are the null and alternative hypotheses respectively, and $\mu_0$ is constant value.

2. Using significant levels $\alpha = 0.01 \ 0r \ 0.05 \ 0r \ 0.10$.

3. We use the statistical test as the following:

    If the number of data is large (bigger than 30), we use
    $$Z = (\frac{\bar{x} - \mu_0}{\sigma})\sqrt{n} \tag{9}$$

    as a statistical test and if the data is small, we use
    $$t = (\frac{\bar{x} - \mu_0}{S})\sqrt{n}. \tag{10}$$

    Where Z, t, and S are the areas under the standard normal distribution curve corresponding the significant level, the student distribution and the standard deviation for small sample respectively.

4. We determine the critical point by comparing $Z_{tabulated}$ and $Z_{calculated}$ as step 3 and the z-table.

5. Decision, we reject or accept the null hypothesis based on step 4.

Table (11) summarizes the properties of the computed parameters as shown in Tables 1,3,5,7 and 9.

**Table (11): the summarized of some cases for above parameters**

| Parameters | No. of cases | Mean $\bar{x}$ | Standard deviation $\sigma$ | Maximum errors (y) |
|---|---|---|---|---|
| $\Delta m$ | 2816 | 2 | 1.75 | 0.065 |
| $m_a$ | 2839 | 8.6 | 1.5 | 0.055 |
| Mass ratio q | 2839 | 0.67 | 0.23 | 0.0085 |
| $\rho''$ | 2005 | 641 | 577.6 | 25.28 |
| a | 2742 | 2561 | 3641 | 136 |

Using the information listed in Table 11 and using Equation (6) we get:
- 95 % C. I. for the mean difference of apparent magnitude ($\mu_{\Delta m}$) is:
$$1.935 \leq \mu_{\Delta m} \leq 2.065$$
- 95 % C. I. for the mean apparent magnitude ($\mu_{m_a}$) is:
$$8.545 \leq \mu_{m_a} \leq 8.655$$
- 95 % C. I. for the mean mass ratio $\mu_q$ is:
$$0.6615 \leq \mu_q \leq 0.6785$$
- 95 % C. I. for the mean projected separation ($\mu_{\rho''}$) is:
$$615.72 \leq \mu_{\rho''} \leq 666.28$$
- 95 % C. I. for the mean linear separation $\mu_a$ is:
$$2425 \leq \mu_a \leq 2697$$

After using Equations (8) and (9), we test the hypothesis for the above parameters using $\alpha = 0.05$ and $Z_{0.025} = 1.96$ $and$ $Z_{0.05} = 2.58$, Table 12 summarize the results.

For the mean of the apparent magnitude difference ($\mu_{\Delta m}$) and if we choose $\mu_0 = 2.5 > 2$, we note that in case of $\mu_{\Delta m} \neq 2.5$ in alternative hypothesis, we reject $H_0$ because the statistical test is located in the rejected region. In case $\mu_{\Delta m} < 2.5$, we reject $H_0$ and if we choose $\mu_{\Delta m} > 2.5$, we accept $H_0$, this means that the mean of population of the difference of apparent magnitude is equal or larger than 2.5, i.e. bigger than 2.

For the mean apparent magnitude $\mu_{m_a}$ and if we choose $\mu_0 = 9$, then after test, we note that, in case $\mu_{m_a} \neq 9$ in alternative hypothesis, we reject $H_0$ because the statistical test is located in the rejected region, if choose $\mu_{m_a} < 9$, we reject $H_0$ and if we choose $\mu_{m_a} > 9$, we accept $H_0$, this means that the mean population of the apparent magnitude is equal to 9 or bigger than 8.6.

For the mean mass ratio $\mu_q$ and if we choose $\mu_0 = 0.7$, then after test, we note that if we choose $\mu_q \neq 0.7$ in alternative hypothesis, we reject $H_0$ because the statistical test is in the rejected region, if we choose $\mu_q < 0.7$, we reject $H_0$ and if choose $\mu_q > 0.7$, we accept $H_0$ this means that the mean population of the mass ratio is equal to 0.7 or bigger than 0.67.

**Table 12: the testing hypothesis of above parameters**

| Parameters | No. of cases | Null and alternative hypothesis | Statistic test (Z) |
|---|---|---|---|
| $\Delta m$ | 2816 | $H_0: \mu_{\Delta m} = 2.5$<br>$H_1: \mu_{\Delta m} \neq 2.5$<br>$or\ \mu > 2.5$<br>$or\ \mu_{\Delta m} < 2.5$ | **15.171** |
| $m_a$ | 2839 | $H_0: \mu_{m_a} = 9,$<br>$H_1: \mu_{m_a} \neq 9$<br>$or\ \mu_{m_a} > 9$<br>$or\ \mu_{m_a} < 9$ | $-14.2$ |
| Mass ratio q | 2839 | $H_0: \mu_q = 0.7,$<br>$H_1: \mu_q \neq 0.7$<br>$or\ \mu_q > 0.7$<br>$or\ \mu_q < 0.7$ | $-6.95$ |
| $\rho''$ | 2005 | $H_0: \mu = 650,$<br>$H_1: \mu \neq 650$<br>$or\ \mu > 450$<br>$or\ \mu < 650$ | $-0.7$ |
| $a$ | 2742 | $H_0: \mu = 2570,$<br>$H_1: \mu \neq 2570$<br>$or\ \mu > 2570$<br>$or\ \mu < 2570$ | $-0.13$ |

For the projected separation, and if we choose $\boldsymbol{\mu_0 = 650}$, we accept H₀ in all three cases, this is wrong, because the standard deviation or the desperation is large, leading to the small test, therefore located in acceptance region. This parameter needs more accurate observations.

For the linear separation, and if we choose $\boldsymbol{\mu_0}$ =2570, also the for the same reason as the projected separation, we accept H₀ in all three cases. The projected and linear separations need separate study.

## 4. Discussion and Conclusion

Some statistical distributions as the frequency distribution of the magnitude difference, the linear separation and the mass ratios are closely related to the evolution not only for the visual binaries but also for the binary systems as a whole. Confidence intervals and testing hypothesis are performed for the five parameters under study. The results reached could be summarized as follows:

- The magnitude difference is concentrated approximately about $\Delta m \simeq 0$ which means that the majority of the systems having components with the same magnitudes.
- The frequency distribution of the mass ratios is concentrated about 0.7 and agree with Salpeter mass function.
- The computed proportion of the degenerate stars do not exceed than 37% of the sample, so, we can consider the sample is bright.
- The linear separation appears to be distributed exponentially, which contradicts with the behavior of close binaries.